\documentclass{Interspeech}

\usepackage{array,booktabs}
\usepackage{color}
\usepackage{cite}
\usepackage{makecell}
\usepackage{multirow}
\usepackage{multicol}
\usepackage{amsmath,graphicx}
\usepackage{amssymb}
\usepackage{booktabs}

\usepackage{arydshln}
\usepackage{pifont}
\newcommand{\xmark}{\ding{55}}%

\interspeechcameraready

\title{Accelerating Diffusion-based Text-to-Speech Model Training\\with Dual Modality Alignment}

\author[affiliation={1}, equalcontribution]{Jeongsoo}{Choi}
\author[affiliation={2,3}, equalcontribution]{Zhikang}{Niu}
\author[affiliation={1}]{Ji-Hoon}{Kim}
\author[affiliation={4}]{Chunhui}{Wang}
\author[affiliation={1}]{Joon Son}{Chung}
\author[affiliation={2,3}]{Xie}{Chen}

\affiliation{School of Electrical Engineering}{KAIST}{South Korea\vskip1pt}
\affiliation{MoE Key Lab of Artificial Intelligence, X-LANCE Lab, School of Computer Science}{\vskip1ptShanghai Jiao Tong University}{China }
\affiliation{}{Shanghai Innovation Institute}{China }
\affiliation{}{Geely}{China}
\email{\{jeongsoo.choi, joonson\}@kaist.ac.kr, \{zhikangniu, chenxie95\}@sjtu.edu.cn}

\begin{document}

\maketitle

\begin{abstract}
The goal of this paper is to optimize the training process of diffusion-based text-to-speech models. While recent studies have achieved remarkable advancements, their training demands substantial time and computational costs, largely due to the implicit guidance of diffusion models in learning complex intermediate representations. To address this, we propose \textbf{A-DMA}, an effective strategy for \textbf{A}ccelerating training with \textbf{D}ual \textbf{M}odality \textbf{A}lignment. Our method introduces a novel alignment pipeline leveraging both text and speech modalities: text-guided alignment, which incorporates contextual representations, and speech-guided alignment, which refines semantic representations. By aligning hidden states with discriminative features, our training scheme reduces the reliance on diffusion models for learning complex representations. Extensive experiments demonstrate that A-DMA doubles the convergence speed while achieving superior performance over baselines.
\end{abstract}

\section{Introduction}
Text-to-Speech (TTS) aims to synthesize natural human speech $x$ from text inputs $y$. Recent advancements, driven by scaling up both data and model size, have enabled the generation of natural speech that closely mimics the voice of any given reference speech $x'$, a capability known as zero-shot TTS. Existing methods for zero-shot TTS can be broadly categorized into two approaches: Autoregressive (AR) ~\cite{wang2023neural,anastassiou2024seed,peng24voicecraft,lajszczak2024base,meng2024autoregressive} and Non-Autoregressive (NAR)~\cite{le2024voicebox,shen2024naturalspeech,eskimez2024e2,chen2024f5,lee2025ditto} models. While AR models have shown exceptional performance in reproducing the acoustic characteristics of reference speaker, they suffer from slow inference speed and error accumulation owing to their structural limitations~\cite{nguyen2025accelerating,han24valler,du2025vall}.

NAR models, primarily built on diffusion networks, address the inherent limitations of AR models, enhancing both inference speed and output quality. For example, VoiceBox adopts a flow-matching algorithm~\cite{lipman2023flow}, outperforming AR-based VALL-E~\cite{wang2023neural} in quality with 20 times faster inference speed. DiTTo-TTS\cite{lee2025ditto} employs a Diffusion Transformer (DiT)~\cite{peebles2023scalable} within latent diffusion models (LDMs), further advancing both speech quality and inference speed. Most recently, E2 TTS\cite{eskimez2024e2} and F5-TTS\cite{chen2024f5} have achieved notable improvements with simpler architectures, demonstrating the drawbacks of external aligners or explicit duration modeling, which are commonly used in prior studies. Despite these advances, training such methods remains time-consuming due to the implicit guidance of diffusion networks in learning complex representations~\cite{hang2023efficient, yao2024fasterdit, yu2025representation}.

In this paper, we present A-DMA, a simple yet effective strategy for Accelerating training through Dual Modality Alignment. Without altering the model architecture, our approach presents two novel alignment pipelines: text-guided alignment and speech-guided alignment, each leveraging modality-specific knowledge from text and speech, respectively. Text-guided alignment utilizes a Connectionist Temporal Classification (CTC)-based compressor to align text information with hidden states, thereby enriching contextual representations within diffusion networks. In parallel, speech-guided alignment incorporates self-supervised speech features into intermediate states through convolutional layers, enhancing the semantic representations within diffusion models.

Given the transformation from text input into the corresponding speech output, we employ text-guided alignment in the earlier diffusion blocks, while speech-guided alignment is applied in the later diffusion blocks. With extensive ablation studies, we validate the design choices of our architecture. Furthermore, experimental results show that our approach achieves state-of-the-art performance on zero-shot TTS benchmarks, confirming the effectiveness of A-DMA from various perspectives. Code and demo samples are available at: \url{https://github.com/ZhikangNiu/A-DMA}

\section{Method}
\begin{figure}[h]
  \centering
  \includegraphics[width=8cm]{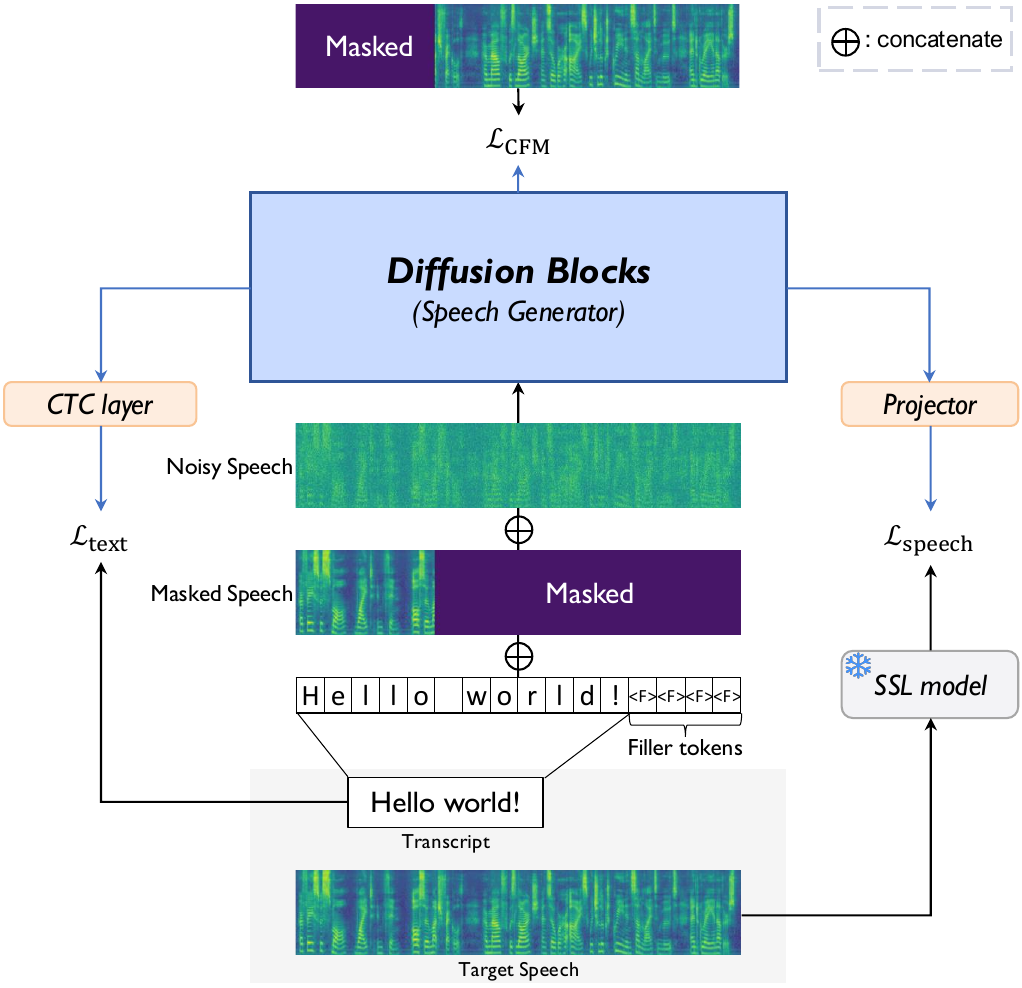}
  \vspace{-0.4cm}
  \caption{Overall training framework. The model uses masked speech, noisy speech, and transcript to predict the masked part of the speech via flow matching. The SSL model and CTC layer are utilized only during training, with the SSL parameters frozen throughout the process.}
  \label{fig:1}
\end{figure}

As illustrated in Figure~\ref{fig:1}, our model introduces an improved training paradigm for a diffusion-based text-to-speech (TTS) system, leveraging dual-modality alignment to accelerate convergence while maintaining a straightforward inference process and high speech fidelity. In this section, we provide a detailed introduction to our proposed method, highlighting its key components and underlying motivations. Section~\ref{method1} introduces our baseline TTS architecture and training scheme. Section~\ref{method2} presents a text-based representation alignment method aimed at improving intelligibility, and Section~\ref{method3} outlines our strategy for enhancing internal speech representations through pre-trained self-supervised learning (SSL) speech models.

\subsection{Diffusion Model for Text-to-Speech}
\label{method1}
Recent advancements in non-autoregressive TTS have established critical foundations for our work. Voicebox\cite{le2024voicebox} is a non-autoregressive flow matching model with a UNet-style Diffusion Transformer, outperforming the previous state-of-the-art TTS model VALL-E\cite{wang2023neural} in intelligibility, speaker similarity, and inference latency. However, its dependency on external duration prediction and forced text alignment introduces pre-processing complexity. Therefore, E2 TTS\cite{eskimez2024e2} eliminates these additional components (duration model, grapheme-to-phoneme conversion, and forced alignment) and converts the input text directly into a character sequence with filler tokens, but E2 TTS suffers from slow convergence and low robustness. To address these limitations, F5-TTS\cite{chen2024f5} introduces three key innovations: 1) A ConvNeXt-based text refinement module to facilitate alignment with speech; 2) An inference-time Sway Sampling strategy, significantly enhancing both the performance and efficiency compared with E2 TTS; 3) Replacing the backbone from U-Net~\cite{ronneberger2015u} style Transformer~\cite{vaswani2017attention} to Diffusion Transformer (DiT)~\cite{peebles2023scalable}.

The flow matching framework aims to construct a probability path $p_t$ between simple distribution $p_0$ (typically Gaussian) and data distribution $q$ through a time-dependent vector field $v_t$, generating flows $\psi_t$ for sampled flow step $t\in[0,1]$. The Conditional Flow Matching (CFM) loss is defined as:
$$\mathcal{L}_{\text{CFM}}=\mathbb{E}_{t,q(x_1),p(x_0)}\Vert v_t(\psi_t(x))-\frac{d}{dt}\psi_t(x) \Vert^2,$$
where $x_1 \sim q$ represents the training audio data and $x_0$ denotes sampled Gaussian noise. We also employ the optimal transport conditional flow matching (OT-CFM) configuration $\psi_t(x) = (1-t)x_0 + tx_1$, simplifying the loss to:
$$\mathcal{L}_{\text{CFM}} = \mathbb{E}_{t, q(x_{1}), p(x_0)} \Vert v_{t}((1-t)x_0+tx_1) - (x_1-x_0) \Vert ^2.$$
In our work, we adopt F5‑TTS as the baseline since it serves as a strong benchmark with state-of-the-art text-to-speech capabilities. The F5-TTS is trained on a text-guided speech-infilling task using optimal transport flow matching, where the model predicts masked speech segments based on noisy speech $\psi_t(x) = (1-t)x_0 + tx_1$, masked speech $x_m = \left(1-m\right)\odot x_1$, and the transcript $y$ with filler tokens, and $m\in\{0,1\}^{F\times{N}}$ represents a binary temporal mask, where $F$ is mel dimension and $N$ is the sequence length.

\subsection{Text Representation Alignment}
\label{method2}
A key requirement in TTS is the generation of intelligible speech. One challenge in TTS is the large discrepancy in sequence lengths between text and speech. To accelerate the learning of correlations between these modalities, we introduce a text representation alignment loss by leveraging an auxiliary CTC~\cite{graves2006connectionist} loss at an intermediate layer of our model. Let $\Phi_i(\psi_t(x), x_m, y)$ denote the hidden representation at the $i$-th Transformer layer, where the inputs include the noisy speech $\psi_t(x)$, the masked speech $x_m$, and the corresponding text embedding $y$. Unlike conventional auxiliary CTC approaches such as InterCTC~\cite{lee2021intermediate, burchi2023audio} primarily used in speech recognition tasks, our approach uniquely applies CTC supervision to the generative model's intermediate states. This early supervision ensures that the model learns the alignment between semantic (text) and acoustic (speech) features at the onset of processing. Formally, the text alignment loss is defined as:
$$\mathcal{L}_{\text{text}} = -\log p_{\text{CTC}}\big(y \,\big|\, \Phi_i(\psi_t(x), x_m, y)\big),$$
where $p_{\text{CTC}}$ is the probability of generating the target text $y$ from the intermediate representation. This loss encourages the model to maintain high fidelity to the input text throughout the transformation process, ultimately leading to more intelligible synthesized speech.

\subsection{Speech Representation Alignment}
\label{method3}
Recent studies~\cite{zhang2024speechtokenizer, ju2024naturalspeech} have demonstrated that incorporating representations from pre-trained self-supervised learning (SSL) speech models, such as HuBERT~\cite{hsu2021hubert}, WavLM~\cite{chen2022wavlm} and W2V-Bert~\cite{chung2021w2v} can significantly enhance the quality of internal representations. Leveraging this insight, we incorporate a speech representation alignment loss into our TTS framework. Specifically, we employ a pre-trained SSL speech model $f$ to extract high-quality representations from the reference speech $x$. To align the temporal and feature dimensions of the intermediate representations from our TTS model with the output of the SSL model, we apply an interpolation layer followed by a 1D convolution layer. This can be formulated as:
$$h_t=\text{Conv1D}(\text{Interp}(\Phi_i(\psi_t(x), x_m, y))).$$
The speech alignment loss is then defined as:
$$\mathcal{L}_{\text{speech}} = -\frac{1}{N} \sum_{n=1}^{N} \cos\big(h_t^{[n]}, \, f(x)^{[n]} \big),$$
where $N$ denotes the total sequence length, and $\text{cos}(\cdot,\cdot)$ computes the cosine similarity between the aligned TTS feature and the SSL feature at each time step. This loss term regularizes the TTS model by encouraging its internal representations to mimic the robust, high-quality features learned by the SSL model.

\subsection{Overall Training Objective}
To jointly optimize our model for both high-quality speech generation and multimodal alignment, we integrate the conditional flow matching loss with the two auxiliary alignment losses. The overall loss function is defined as follows:
$$\mathcal{L}_\text{total}=\mathcal{L}_\text{CFM} + \lambda_\text{text}\mathcal{L}_\text{text} + \lambda_\text{speech}\mathcal{L}_\text{speech},$$
where $\lambda_\text{text}$ and $\lambda_\text{speech}$ are hyperparameters that control the relative importance of the text and speech alignment losses. In our experiments, they are empirically set to 0.1 and 1, respectively, to balance their contributions. By optimizing this joint objective, the model learns to synthesize speech that is not only natural and expressive but also well-aligned with the input text and the reference speaker’s characteristics.

\section{Experiments}

\begin{table}[t]
  \renewcommand{\arraystretch}{1.3}
  \renewcommand{\tabcolsep}{2.3mm}
  \centering
  \caption{Experiments on LibriSpeech-PC test-clean dataset. The best results among low-resource setting are marked in \textbf{bold}.}
  \vspace{-0.2cm}
  \label{tab:table1}
  \resizebox{0.999\linewidth}{!}{
  \begin{tabular}{ l c c c c }
    \Xhline{3\arrayrulewidth}
    \textbf{Method} & \# Param. & Data & \textbf{WER}\((\%)\)$\downarrow$ & \textbf{SIM}$\uparrow$ \\
    \hline
    Ground Truth & & & 2.23 & 0.69 \\
    \hdashline
    \multicolumn{5}{c}{\itshape High-resource} \\
    CosyVoice~\cite{du2024cosyvoice} & 300M & 170kh & 3.59 & 0.66 \\
    FireRedTTS~\cite{guo2024fireredtts} & 580M & 248kh & 2.69 & 0.47 \\
    E2 TTS~\cite{eskimez2024e2} & 333M & 100kh & 2.95 & 0.69 \\
    F5-TTS~\cite{chen2024f5} & 336M & 100kh & 2.42 & 0.66 \\
    \hdashline
    \multicolumn{5}{c}{\itshape Low-resource} \\
    USLM~\cite{zhang2024speechtokenizer} & 361M & 0.6kh & 6.11 & 0.43 \\
    E2 TTS~\cite{eskimez2024e2} & 157M & 0.6kh & 3.10 & 0.63 \\
    ~~~+ \textbf{A-DMA (Ours)} & 157M & 0.6kh & 2.25 & \textbf{0.65} \\
    F5-TTS~\cite{chen2024f5} & 159M & 0.6kh & 2.68 & 0.60 \\
    ~~~+ \textbf{A-DMA (Ours)} & 159M & 0.6kh & \textbf{1.97} & 0.62 \\
    \Xhline{3\arrayrulewidth}
  \end{tabular}}
  \vspace{-0.2cm}
\end{table}


\subsection{Dataset}
We train our model on the LibriTTS~\cite{zen2019libritts} dataset, a multi-speaker English corpus containing approximately 585 hours of read speech sampled at 24~kHz. The dataset is derived from the LibriSpeech~\cite{panayotov2015librispeech} corpus and curated specifically for TTS research. Utterances with significant background noise are excluded to ensure high-quality data for training.

\subsection{Preprocessing}
For text, we use character sequences for both input and text-guided alignment. For speech, we use audio with 24~kHz sampling rate and extract mel-spectrograms with 100 mel filter banks, a window size of 1024, and a hop size of 256, resulting in approximately 94~Hz. Since the mel-spectrogram sequence is generally much longer than the character sequence, filler tokens are concatenated to the character sequence to match the length of the mel-spectrogram. For speech-guided alignment, we resample the audio to 16~kHz before feeding it into the speech SSL model.

\subsection{Model Setup}
Our baseline model is F5-TTS Small, which consists of 18 layers, 12 attention heads, and a 768-dimensional embedding for the DiT block; and 4 layers with a 512-dimensional embedding for the ConvNeXt V2~\cite{woo2023convnext} module, totaling 159M parameters. We train the model with a total batch size of approximately 1k seconds, using the AdamW~\cite{loshchilov2019decoupled} optimizer with a learning rate of \(7.5 \times 10^{-5}\). The learning rate is scheduled with a warmup phase of 20,000 updates, followed by linear decay. Training is conducted on 4 RTX A6000 GPUs.

For inference, we apply Exponential Moving Average (EMA) weights and the sway sampling strategy with a coefficient of -1, utilizing the Euler ODE solver for F5-TTS, and the midpoint method for E2-TTS, as described in \cite{chen2024f5}. To synthesize audio from the generated mel-spectrograms, we use the pre-trained vocoder Vocos~\cite{siuzdak2024vocos}, which was trained on LibriTTS for 1.2M steps.

\begin{figure}[ht]
    \centering
    \includegraphics[width=8cm]{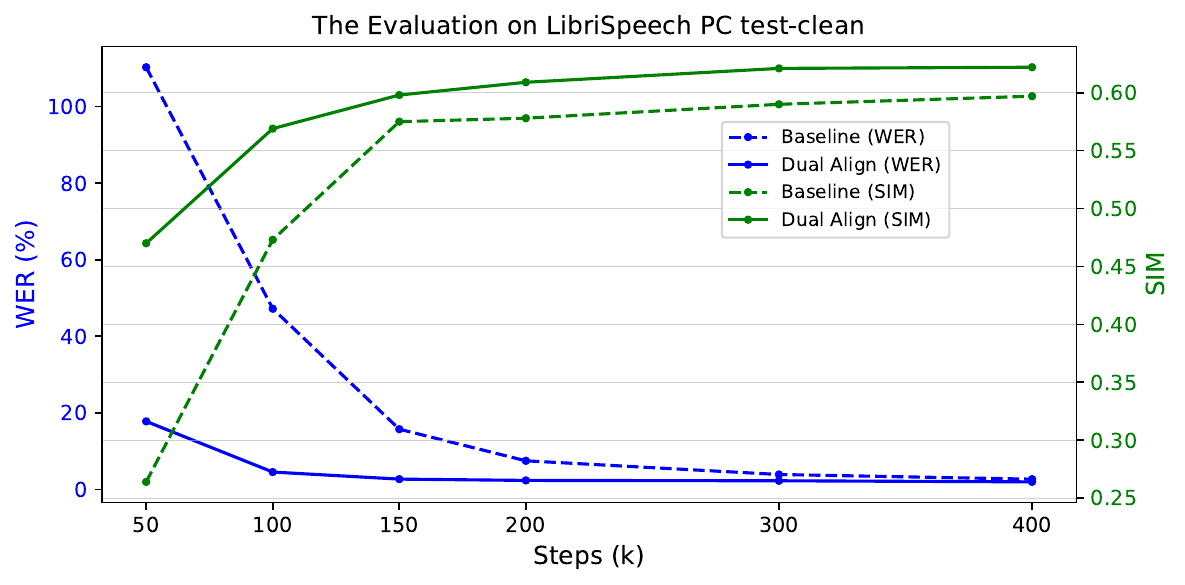}
    \vspace{-0.5cm}
    \caption{The WER and SIM results showing the effectiveness of the proposed dual modality alignment.}
    \label{fig:2}
\end{figure}

\subsection{Evaluation}
We follow the evaluation setup used in F5-TTS to assess our model on the LibriSpeech-PC~\cite{meister2023librispeech} test set. This test set consists of 2.2 hours of data from the LibriSpeech~\cite{panayotov2015librispeech} test-clean subset, comprising 1,127 samples, each ranging from 4 to 10 seconds in length. It is specifically designed to facilitate comparison across different models in the research community and is publicly available for benchmarking. Specifically, we assess its ability to generate speech that aligns with the content of the provided text and the speaker characteristics of the given speech prompt. We perform the cross-sentence generation task using two utterances of the same speaker, where one serves as the speech prompt and the other is the target speech to be synthesized.
The evaluation is conducted using the following three metrics:

\noindent \textbf{Word Error Rate (WER)} measures the intelligibility of synthesized speech by comparing its transcription to the ground-truth text. We employ Whisper-large-V3~\cite{radford2023robust} for automatic transcription and compute WER accordingly.

\noindent \textbf{Speaker Similarity (SIM)} quantifies the resemblance between the synthesized and ground-truth speech in terms of speaker identity. We use the WavLM-large-based~\cite{chen2022wavlm} speaker verification model to extract speaker embeddings and compute their cosine similarity between synthesized and ground truth speeches.

\noindent \textbf{UTMOS}~\cite{saeki2022utmos} is an objective metric for assessing the naturalness of synthesized speech. It employs an open-source MOS prediction model to estimate audio quality without requiring reference recordings or labels. While not an absolute measure, UTMOS provides a practical and efficient way to gauge the perceived naturalness of synthetic speech.

\begin{table}[ht]
  \renewcommand{\arraystretch}{1.4}
  \renewcommand{\tabcolsep}{1mm}
  \centering
  \caption{Ablation study of modality aligners on LibriSpeech-PC test-clean dataset. All of the models are trained on LibriTTS for about 100 epochs. Here, ``\textit{speech align}'' uses a HuBERT-large alignment, and ``\textit{text align}'' uses a CTC  alignment.}
  \vspace{-0.2cm}
  \label{tab:table2}
  \resizebox{0.999\linewidth}{!}{
  \begin{tabular}{l c c c c c}
    \Xhline{3\arrayrulewidth}
    Method & $\mathcal{L}_{\text{text}}$ & $\mathcal{L}_{\text{speech}}$ & \textbf{WER}(\%)$\downarrow$ & \textbf{UTMOS}$\uparrow$ & \textbf{SIM}$\uparrow$ \\
    \hline
    Baseline & \xmark & \xmark & 7.474 & 4.040 & 0.578 \\
    \hdashline
    \multirow{5}{*}{+ \textbf{text align}}
    & 2nd layer & \xmark & 7.188 & 4.053 & 0.591 \\
    & 4th layer & \xmark & 4.558 & 4.052 & 0.592 \\
    & 8th layer & \xmark & 2.602 & 4.076 & 0.586 \\
    & 12th layer & \xmark & 2.483 & 4.079 & 0.589 \\
    & 16th layer & \xmark & 2.965 & 4.033 & 0.587 \\
    \hdashline
    \multirow{5}{*}{+ \textbf{speech align}}
    & \xmark & 2nd layer & 8.288 & 4.056 & 0.591 \\
    & \xmark & 4th layer & 7.421 & 4.091 & 0.592 \\
    & \xmark & 8th layer & 4.900 & 4.121 & 0.606 \\
    & \xmark & 12th layer & 3.521 & 4.089 & \textbf{0.609} \\
    & \xmark & 16th layer & 4.470 & 4.012 & 0.600 \\
    \hdashline
    \multirow{4}{*}{+ \textbf{dual align}}
    & 8th layer  & 8th layer & 3.063 & 4.087 & 0.606 \\
    & 12th layer & 12th layer & 2.688 & 4.076 & 0.604 \\
    & 8th layer  & 12th layer & 2.353 & 4.054 & \textbf{0.609} \\
    & 8th layer & 16th layer & \textbf{2.226} & 4.024 & 0.600 \\
    \Xhline{3\arrayrulewidth}
  \end{tabular}}
  \vspace{-0.3cm}
\end{table}


\section{Results}
\subsection{Quantitative Comparison}
Table~\ref{tab:table1} shows the evaluation results of previous methods and our proposed approach \textbf{A-DMA} on LibriSpeech-PC~\cite{meister2023librispeech} test-clean dataset. As shown in high-resource data setting, E2 TTS and F5-TTS which are diffusion-based NAR models demonstrate comparable performance to state-of-the-art AR models such as CosyVoice~\cite{du2024cosyvoice} and FireRedTTS~\cite{guo2024fireredtts}. To assess the effectiveness of A-DMA, we apply it to E2 TTS and F5-TTS while keeping all training settings identical, with only our method added. The results demonstrate that our method achieves superior performance in both WER and SIM metrics than existing methods including USLM~\cite{zhang2024speechtokenizer}. Notably, our results highlight the effectiveness of the proposed dual alignment method in enhancing the alignment between input text and generated speech, even with limited training data. Importantly, our approach does not modify the inference pipeline, ensuring that the inference speed remains unchanged while achieving improved performance.

To further analyze the effectiveness of our approach, Figure~\ref{fig:2} illustrates the performance trends across training steps. The results indicate that our method significantly accelerates the alignment process between the input text and generated speech while maintaining higher performance upon convergence, demonstrating its robustness and efficiency in low-resource scenarios.

\subsection{Layer-wise Analysis}
We conduct extensive layer-wise experiments to investigate the impact of alignment at different layers as shown in Table~\ref{tab:table2}. Our findings indicate that each modality achieves optimal performance when aligned at the 12th layer. As expected, text alignment significantly enhances intelligibility (WER), while speech alignment improves speaker similarity. Applying alignment at the early layers for both modalities does not lead to substantial improvements, suggesting that these layers are insufficient to effectively align with discriminative representations.

Furthermore, we compare alignment at the 8th and 16th layers and observe that applying text guidance at a lower layer and speech guidance at a higher layer yields superior performance. These results suggest that optimal alignment occurs at different depths for each modality. To further investigate, we apply our proposed dual alignment approach. We find that aligning the same 12th layer for both modalities leads to suboptimal performance, whereas single modality alignment achieves better results. This is reasonable given the inherent representation gap between text and speech features, causing a trade-off when aligning both simultaneously. To address this, we explore aligning text at a lower layer than speech alignment, finding that this strategy significantly improves performance, outperforming both single-modality alignment and the baseline.

\begin{table}[ht]
  \renewcommand{\arraystretch}{1.5}
  \renewcommand{\tabcolsep}{1.0mm}
  \centering
  \caption{Ablation study of speech SSL model and speech align equation after 100 epochs trained on LibriTTS dataset - $\mathcal{L}_{speech}$ is applied to 12th layer. ``Avg." refers to the average of the intermediate outputs of the SSL model, while ``Last" refers to using only the output of the last layer of the SSL model.
  }
  \vspace{-0.2cm}
  \label{tab:table3}
  \resizebox{0.999\linewidth}{!}{
  \begin{tabular}{ c c c c c c }
    \Xhline{3\arrayrulewidth}
    SSL model & Output & $\mathcal{L}_{\text{speech}}$ & \textbf{WER}\((\%)\)$\downarrow$ & \textbf{UTMOS}$\uparrow$ & \textbf{SIM}$\uparrow$ \\
    \hline
    \multicolumn{2}{c}{Baseline} & \xmark & 7.474 & 4.040 & 0.578 \\
    \hdashline
    HuBERT & Last & $|\Phi_n-f(x)|$ & 6.386 & 4.068 & 0.605 \\
    HuBERT & Last & $-\log(\sigma(\cos(\Phi_n, f(x))))$ & 5.104 & 4.056 & 0.606 \\
    HuBERT & Last & $-\cos(\Phi_n, f(x))$ & \textbf{3.521} & 4.089 & 0.609 \\
    HuBERT & Avg. & $-\cos(\Phi_n, f(x))$ & 4.716 & 4.070 & \textbf{0.618} \\
    WavLM & Last & $-\cos(\Phi_n, f(x))$ & 3.992 & 4.100 & 0.612  \\
    \Xhline{3\arrayrulewidth}
  \end{tabular}}
  \vspace{-0.3cm}
\end{table}


Notably, our method maintains a high UTMOS score, demonstrating that the improvements in alignment do not come at the cost of naturalness. The UTMOS scores remain comparable with the baseline, ensuring that our dual alignment approach preserves speech quality and naturalness while achieving enhanced intelligibility and speaker similarity.

\subsection{Ablation Study}
Table~\ref{tab:table3} presents an ablation study on the speech alignment method. 
We experiment with various losses and determine that the negative cosine similarity loss (line \#3, \#4, and \#5) yields the best results compared to L1 (line \#1) and the negative cosine similarity loss with sigmoid activation (line \#2). When we use the average layer output of the SSL speech model, we find it improves speaker similarity (SIM) slightly, but degrades the WER compared to using the last layer alone. This trade-off suggests that averaging across all layers may introduce information redundancy, which can bias the model by incorporating less relevant or noisy features from earlier layers. For all other experiments, we adopt the last layer of HuBERT with negative cosine similarity loss.

\section{Conclusion}
In this paper, we propose A-DMA, a novel modality alignment strategy that effectively guides diffusion-based text-to-speech models by leveraging discriminative representations. This approach significantly accelerates model convergence while enhancing the quality of generated speech. Experimental results validate that our approach outperforms the baseline system in both training efficiency and speech quality while maintaining a straightforward inference pipeline. We further validate the effectiveness of each proposed component through comprehensive layer-wise analysis and ablation studies. We believe that these promising findings pave the way for future research in advanced modality alignment techniques and further optimization of text-to-speech systems.

\section{Limitation}
Although our extensive layer-wise experiments demonstrate that the proposed dual modality alignment leads to superior generation performance with faster convergence, this work does not target model efficiency. In future work, we plan to further optimize the framework by reducing model complexity and accelerating inference.

\section{Acknowledgements}
J. Choi, J.-H. Kim, and J. S. Chung  were supported by IITP grant funded by the Korea government (MSIT) (RS-2024-00457882, National AI Research Lab Project). Z. Niu and X. Chen were supported by the National Natural Science Foundation of China (No. U23B2018 and No. 62206171), Shanghai Municipal Science and Technology Major Project under Grant 2021SHZDZX0102.

\bibliographystyle{IEEEtran}
\bibliography{mybib}

\end{document}